# Volatility-Volume Order Slicing via Statistical Analysis

Ritwika Chattopadhyay, Abhishek Malichkar, Zhixuan Ren, Xinyue Zhang

## Abstract

This paper addresses the challenges faced in large-volume trading, where executing substantial orders can result in significant market impact and slippage. To mitigate these effects, this study proposes a volatility-volume-based order slicing strategy that leverages Exponential Weighted Moving Average and Markov Chain Monte Carlo simulations. These methods are used to dynamically estimate future trading volumes and price ranges, enabling traders to adapt their strategies by segmenting order execution sizes based on these predictions. Results show that the proposed approach improves trade execution efficiency, reduces market impact, and offers a more adaptive solution for volatile market conditions. The findings have practical implications for large-volume trading, providing a foundation for further research into adaptive execution strategies.

## 1. Introduction

Traders in standard trades aim to achieve "best execution," which refers to executing orders at the most favorable price possible, considering factors like liquidity, timing, and market conditions. Their goal is to minimize transaction costs and market impact, ensuring the trade is executed as close as possible to the expected price. However, in large-volume trading, the situation becomes more complex. Executing large orders often creates significant market pressure, which can disrupt the normal price dynamics. As the size of the order increases, maintaining the desired price becomes more difficult, and the order itself can move the market. These shifts frequently lead to slippage, where trades fill at prices different from expectations due to limited liquidity or sudden price movements. Therefore, effectively managing these risks is crucial for high-volume traders to sustain profitability and minimize potential losses in volatile markets.

To reduce market risk, we utilize a volatility-volume-based order slicing strategy, incorporating Exponentially Weighted Moving Average (EWMA) and Markov Chain Monte Carlo (MCMC) simulation. The EWMA model is used to forecast short-term price and volume trends by giving more weight to recent data, while MCMC simulates future price ranges and volumes by sampling from their posterior distributions. Together, these techniques provide dynamic estimates of future volume and price ranges, enabling traders to adjust their order-slicing strategy, minimizing market impact, and improving the overall efficiency of trade execution.

The report is structured as follows: [Section 2](#) describes the data collection process and preprocessing steps. [Section 3](#) outlines the methodology employed in the study. [Section 4](#) presents the analysis and results. Finally, Section 5 provides conclusions, along with recommendations and implications for future research.

## 2. Data

### 2.1 Data Collection

The data for this study was sourced from Yahoo Finance (Yahoo Finance, 2024) and was chosen for its reliability and API support for intraday stock data retrieval. The API allows access to high-frequency financial data, enabling detailed analysis of market behavior. For this study, we collected one month of intraday data for Tesla Inc. (Ticker: TSLA).

The dataset encompasses regular trading hours from 09:30:00 to 15:59:00, for all trading days within the selected month. This focus on regular trading hours ensures the exclusion of pre-market and after-hours sessions. The dataset includes the following columns that can be referred to in Appendix 1:

*Table 1: Data Column Description*

| Column Name | Definition |
| --- | --- |
| DateTime | The timestamp for each minute interval. |
| Open Price | The price of the stock at the start of each minute interval. |
| High Price | The highest price is recorded within each minute interval. |
| Low Price | The lowest price is recorded during each minute interval. |
| Closing Price | The price of the stock at the end of each minute interval. |
| Adjusted Closing Price | The adjusted price of the stock at the end of each minute interval. |
| Volume | The total number of shares traded within the corresponding minute interval. |

This structured data captures Tesla's intraday trading dynamics, offering a detailed view of price movements and trading activity over time. By leveraging these fields, we aim to quantify and analyze volatility and volume patterns at a granular level. This dataset forms the foundation of our investigation, enabling a detailed exploration of the interplay between intraday volatility and trading volume, and its implications for order slicing strategies in high-frequency trading environments.

## 2.2 Data Preprocessing

To enhance the clarity and stability of the analysis, the collected one-minute interval data underwent two levels of aggregation: 5-minute intervals and daily summaries. This preprocessing step reduces the noise inherent in high-frequency data while retaining essential information for analyzing volatility and volume.

The first step involved grouping the one-minute intraday data into 5-minute intervals as shown in [Appendix 2](). While one-minute data provides detailed insights into short-term price movements, it is often highly noisy due to fluctuations and outliers, which can obscure underlying trends. By aggregating the data into 5-minute intervals, we smooth out these irregularities, making the data more stable for statistical analysis. Additionally, this transformation improves the reliability of the probability density function (PDF) by increasing the volume of data points within each interval, which reduces sparsity and enhances the representation of price and volume distributions, which will be explained in [3.1 Distribution Identification]().

*Table 2: 5-Minute Data Aggregation Column Description*

| Column Name | Definition |
| --- | --- |
| DateTime | The timestamp for the 5-minute interval. |
| Open Price | The first price is recorded within the 5-minute interval. |
| High Price | The highest price is recorded within the 5-minute interval. |
| Low Price | The lowest price is recorded during the 5-minute interval. |
| Closing Price | The price of the stock at the end of the 5-minute interval. |
| Range | The difference between the High and Low prices, represents volatility within the interval. |
| Range_Up | Decomposed upward price movement within the interval. |
| Range_Down | Decomposed downward price movement within the interval |

Following the 5-minute aggregation, the data was further grouped at the daily level. This step aggregates all trading activity within regular market hours for each day into a single summary, capturing overall trends in price movements and trading volume. The relevance of the daily-level data will be further elaborated in [3.4 Mapping of EWMA to Expected Range Values]().

## 3. Methodology

This methodology section illustrates a comprehensive framework for analyzing the relationship between trading volumes and price movements. The process begins by identifying the conditional distribution of trading volumes with respect to price ranges, leveraging Maximum Likelihood Estimation to characterize a log-normal distribution. The Metropolis-Hastings algorithm is then employed to estimate expected price ranges and corresponding trade volumes, ensuring robust probabilistic predictions. To enhance forecast accuracy, the Exponentially Weighted Moving Average is applied to predict price movement ranges, with a mapping process ensuring consistency between different time granularities. These steps collectively enable the estimation of trade volumes and price ranges, supporting strategies that minimize market impact and optimize order execution.

### 3.1 Distribution Identification

This section aims to determine the conditional distribution of trading volume with respect to price ranges. First, the datasets from Sections 2.1 and 2.2 are merged to create a new dataset containing three columns: Datetime, Volume, and Range, as presented in Appendix 3. Upon examining this dataset, it is observed that the trading volume exhibits a right-skewed pattern, which is indicative of a log-normal distribution, as shown below. Based on this observation, the price range is divided into several bins, with the number of bins ranging from 5 to 10. For each bin configuration, the log-normal distribution parameters—location($\mu$), scale($\theta$), and shape($\sigma$)—are estimated for the volume data using Maximum Likelihood Estimation (MLE). These parameters are systematically recorded for further analysis.

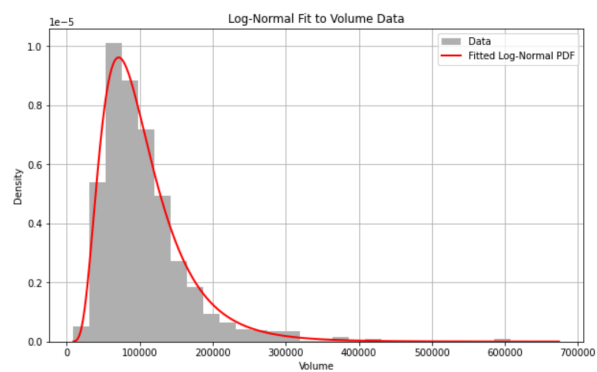

*Graphic 1: Distribution of Trading Volume*

This step serves as the foundation for subsequent methods, specifically referred to in Section 3.2.2 Detailed Methodology. The optimal bin configuration is selected by minimizing the discrepancy between the expected and observed trading volumes across the price ranges, ensuring reliable input for the following modeling and estimation processes.

## 3.2 Metropolis-Hastings Algorithm for Range Value

The primary objective of this methodology is to estimate expected ranges using a Markov Chain Monte Carlo (MCMC) method, specifically the Metropolis-Hastings algorithm(Hachicha & Hachicha, 2020). This approach leverages the relationship between high-low price ranges and traded volumes to provide robust probabilistic estimates, ultimately supporting strategies for order slicing and execution with minimal market impact. Below, we outline the detailed steps that led to the development and implementation of this model.

The dataset comprises intraday high-low price ranges and traded volumes. Key steps involved:

1. **Range Calculation**: The high-low price ranges were computed for each time interval.
2. **Distribution Analysis**: The distribution of the range values was analyzed. Visualizations revealed a skewed distribution with heavy tails, suggesting that a log-normal distribution was the best fit.
3. **Binning**: The range values were divided into bins to group ranges into meaningful intervals. Each bin was characterized by specific parameters (shape, location, and scale) of the log-normal distribution, obtained through fitting the data in that bin.
4. **Parameter Extraction**: For each bin, the parameters of the fitted log-normal distribution were stored, including the shape parameter ($\sigma$), location ($\mu$), and scale ($\theta$). These parameters represent the statistical characteristics of the ranges and corresponding volumes in the dataset.

### 3.2.1 Overview of Metropolis-Hastings Algorithm

The Metropolis-Hastings algorithm was chosen as the MCMC method to simulate the probability distribution of range values conditioned on volumes. The algorithm generates samples iteratively, balancing exploration and convergence to the target distribution.

The posterior probability to be estimated is: (Lognormal prior) x likelihood, where prior is the stationary prior distribution of the range values, assumed to follow a log-normal distribution based on prior analysis. Likelihood is observing a given volume given a range, derived from the log-normal distribution parameters for the corresponding bin.

### 3.2.2 Detailed Methodology

The stationary prior distribution of range values was defined as a log-normal distribution parameterized by:
- Shape ($\sigma$): Variance of the stationary distribution.
- Location ($\mu$): Mean shift of the distribution.

- Scale (θ): Scale parameter of the distribution.

This distribution represents the prior belief about the range values before observing specific volumes. These parameters are computed from historical data for each bin range. The probability density function (PDF) of the log-normal distribution is given as

$$f(x; s, \mu, \sigma) = \frac{1}{x * s\sqrt{2\pi}} * exp(-\frac{(ln(x)-u)^2}{2s^2}), x > 0$$

This PDF provides the likelihood of observing specific range values given the parameters of the distribution. The MH algorithm begins by sampling an initial range value, Rprev, from a stationary prior distribution defined by the log-normal parameters (s,μ,σ) of the overall range dataset. At each iteration, a new range value, Rproposal, is proposed by sampling from the same stationary distribution. The sampled value is evaluated by comparing its likelihood with that of the previous range value. For a given bin range, the likelihood of a range value R is calculated as follows

$$P(Volume|R) = f(Volume, s, \mu, \sigma)$$

Where f is the log-normal PDF, and the parameters s,μ,σ are specific to the bin into which the proposed range value falls. The algorithm identifies the bin range of the proposed value Rproposal by checking which range interval the value falls into. The closest observed range value within that bin range is then identified, and the volume associated with this observed range value is retrieved. This observed volume is used to calculate the likelihood of the proposed value. The likelihood ratio, which determines whether the proposal is accepted, is given by

$$\text{Acceptance Criteria} = min(1, \frac{P(Volume | R_{proposal})}{P(Volume | R_{prev})})$$

If the acceptance criterion is met, the proposed range value is accepted, and RproposalRproposal becomes the new RprevRprev. Otherwise, the previous range value Rprev is retained. This process iterates over many steps, generating a Markov chain of sampled range values that converge to the posterior distribution of P(Range│Volume)P(Range│Volume).

A key refinement in the algorithm ensures that the proposed range value is compared with the closest observed range value in the dataset within the corresponding bin range. This ensures that the sampled range values are compared in realistic market scenarios. The algorithm outputs a sequence of accepted range values that represent the posterior distribution and reduces the market impact which will be described in the later section.

### 3.3 Identification of Price Movement Range Using EWMA

As discussed in the previous section, different bin groupings and their corresponding expected values have been identified. To determine the most appropriate expected value, referred to as the predicted range for the next day's price movement, the Exponentially Weighted Moving Average (EWMA) technique is employed.

EWMA is a statistical method that processes time-series data by assigning exponentially decreasing weights to older records, giving greater importance to more recent data points (Adewuyi, 2016). This approach ensures that the most recent observations heavily influence the prediction while older observations contribute progressively less. The EWMA can be expressed using the following equation:

$$S_t = \lambda X_t + (1 - \lambda) S_{t-1}$$

where:

- $S_t$ is the EWMA at time $t$,
- $X_t$ is the observed value at time $t$,
- $\lambda$ is the smoothing factor ($0 < \lambda < 1$),
- $S_{t-1}$ is the EWMA at time $t - 1$

Lambda ($\lambda$) here is defined as the smoothing factor of EWMA. It is more intuitive to define $\lambda$ in terms of the number of periods we want to look back at, also known as the half-life ($h$).

$$\lambda = 1 - e^{-ln(2)/h}$$

In this study, we consider a look-back period corresponding to one trading week. For instance, if the current day is Wednesday of Week 3 and we are predicting the price movement for Thursday of Week 3, the look-back period will include the following days: Thursday and Friday of Week 2, as well as Monday, Tuesday, and Wednesday of Week 3. This approach ensures that the predicted range is informed by the most recent trends while accounting for temporal dynamics in the dataset.

**3.4 Mapping of EWMA to Expected Range Values**

The EWMA values are calculated based on data aggregated at a weekly level. To provide clarity, we use daily price ranges derived from the highest and lowest prices recorded throughout the day. This aggregation ensures that the EWMA captures broader market trends without being overly sensitive to short-term fluctuations. [Appendix 4:](#) shows a reference snippet illustrating the data grouped at a daily level.

Once the EWMA metric is computed, the predicted range is determined by selecting the expected range from the previous step that is closest to the EWMA value from the relevant look-back period. However, the expected range values from the previous step are derived by grouping data points at 5-minute intervals.

To ensure consistency between the EWMA range (computed at the daily level) and the expected range (calculated at 5-minute intervals), it is necessary to map the daily EWMA values to a corresponding range at the 5-minute interval level. This mapping is achieved through the application of a scaling factor. The scaling factor that maps the expected range to the EWMA range can be expressed as:

$$ASF = \frac{1}{k}\sum_{i=1}^{k} \frac{R_i}{\overline{R_i}}$$

where:

- $R_i$ is the EWMA range for day $i$,
- $\overline{R_i}$ is the average 5-minute range for day $i$,
- $k$ is the number of days looking back at (in this case 5)

By applying the scaling factor, that is dividing the EWMA of our look-back period by the ASF value, we can compare the modified EWMA range with the expected range values from the previous step and select the closest match, ensuring consistency and accuracy in predictions.

**3.5 Metropolis-Hastings Algorithm for Volume for Specific Range**

To estimate the potential trade volume for the trading day, the methodology based on the Metropolis-Hastings (MH) algorithm builds upon the expected range value derived earlier.

This approach seeks to identify a set of possible volume values that maximizes the likelihood of observing the desired range (the output derived from the previous section) while minimizing market impact. The key idea is to treat the expected range as a hard constraint, ensuring that the probability of observing the range is independent of the trade volume. This independence ensures that the selected trade volumes are aligned with minimizing market impact without compromising the likelihood of achieving the expected range.

The methodology assumes that the range value and volume are independent, expressed as $P(R|V)=P(R)$. This independence allows us to treat the expected range as fixed while sampling volumes iteratively. The algorithm aims to ensure that the likelihood of the range remains unaffected by the volume being traded. This assumption forms the foundation for the volume sampling process, as it helps prioritize trade volumes that are less likely to disrupt market dynamics while maintaining the desired range.

The MH algorithm is then applied to sample potential volume values from a posterior distribution conditioned on the expected range. The process starts with initializing a starting volume value ($V_{prev}$), which is sampled from a prior log-normal distribution. The parameters of this prior distribution are derived from historical trade data, ensuring that the starting point reflects realistic trading scenarios. The expected range value, obtained from the previous part of the MH simulation, is treated as a fixed input throughout this process.

At each iteration of the algorithm, a new volume value, *V* proposal, is proposed by sampling from the same log-normal prior distribution. The likelihood of observing the expected range, given both the proposed and the previous volume values, is calculated using the log-normal probability density function. Since the expected range is treated as independent of the volume, the likelihood calculation simplifies evaluating the probability of the proposed and previous volume values under the log-normal prior. The acceptance criterion for the proposal is determined by the ratio of these probabilities. A uniformly random number is generated, and the proposed volume is accepted if this number is less than or equal to the calculated acceptance probability.

Once the sampling is complete, the collected volume values are analyzed to identify trends and insights. A log-normal distribution is fitted to the sampled volumes to provide a concise summary of the results. This fitted distribution captures the characteristics of the posterior distribution and serves as the basis for order execution strategies.

## 4. Results & Discussion

After identifying the expected range and the volume to be traded for the next day, our objective is to evaluate the accuracy of our predictions against the actual trading data. Before delving into the metrics, it is important to outline the timeline and methodology used for evaluation.

The proposed evaluation framework follows a rolling window approach. Given the availability of intraday information, we utilize the last five trading days' data at any given point to generate probability distributions. These distributions are simulated using the **Metropolis-Hastings algorithm**, enabling the generation of expected ranges and trading volumes for the subsequent day.

For instance, the information from the first five days is used to predict the trading volume and the expected range (at 5-minute intervals) for the 6th day. To bring the predicted range to a daily level scale, we adopt the **Adjusted Scaling Factor (ASF)**, computed from the past five days.

On the 6th day, the prediction is evaluated using the actual trading data. Specifically:

1. **Predicted Trading Volume:** Compared to the average trading volume recorded on day 6.
2. **Predicted Range:** Validated against the **EWMA range value** calculated for day 6.

The graphic below shows how the testing is done using a rolling window without allowing any form of data leakage:

*Graphic 2:Rolling Window Testing*

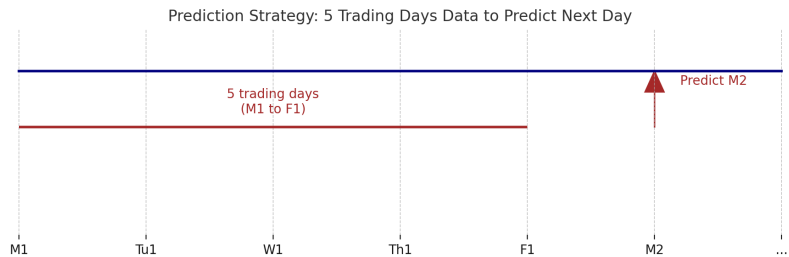

**M1 to F1 (5 trading days)**: These represent the past 5 trading days, from Monday (M1) to Friday (F1). The data for these 5 days are used to calculate the prediction for the next trading day. Each day is denoted as a separate entity (M1, Tu1, W1, Th1, F1), corresponding to a week's Monday to Friday.

**Predict M2**: Using the data from M1 to F1, the model predicts the range and volume for the next Monday (M2).

The rolling window strategy ensures that after predicting M2, the timeline shifts forward, and the next prediction will be made using the most recent 5 trading days (now starting from Tu1 and ending with M2, and so on). Every step starting from identifying the distribution is again initiated from scratch, preventing any leakage. This rolling evaluation process ensures that our model consistently adapts to recent market dynamics, enabling robust predictions and reliable performance assessment.

The volume predicted versus the actual average volume traded has been plotted in the graph below:

*Graphic 3:Comparison of Predicted & Actual Volume*

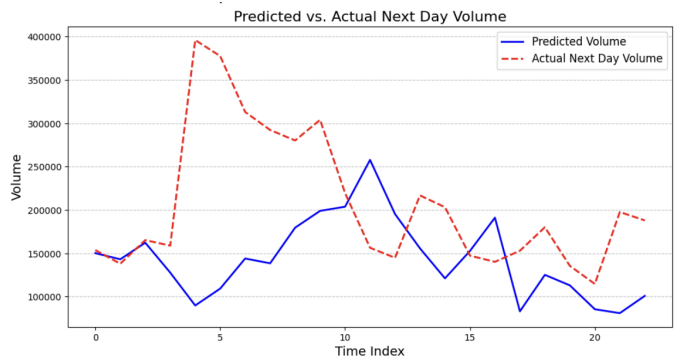

The primary evaluation metric used in this study is Mean Absolute Percentage Error (MAPE), which is an important statistical evaluation metric for forecasting or prediction models. It is given by the ratio as follows:

$$MAPE = \frac{100}{n} \sum_{i=1}^{n} \frac{|A_i - F_i|}{A_i}$$

where:

- $A_i$ is the actual value for the day $i$,
- $F_i$ is the predicted value for the day $i$,
- $n$ is the number of observations being tested.

The final results indicate that the MAPE for the predicted volume is around 26% and predicted range is around 35%, highlighting the model's forecasting accuracy.

*Table 3:Error Metric*

| Metric | Value |
|---|---|
| Average MAPE for Volume | 26.23% |
| Average MAPE for Range | 35.88% |

While a 35% MAPE may not align with the best industry standards, it remains within acceptable limits given the timeframe under consideration. Another important factor to explore is the selection of stocks and the timeframe used to test and evaluate our model. Following the 45th Presidential Election, Tesla stock exhibited notable fluctuations in both price and trading volume as be seen in Graphics 4 and 5. Since our model relies on historical data, the recency of this change suggests that the study can be further strengthened by extending and testing the strategy on other stocks to achieve more stabilized and robust results.

*Graphic 4:Tesla Daily Price Fluctuation*

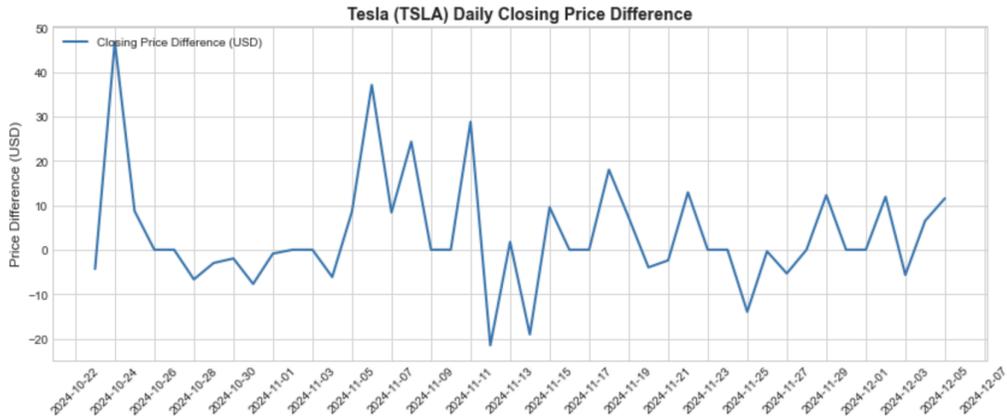

*Graphic 5:Tesla Daily Volume Fluctuation*

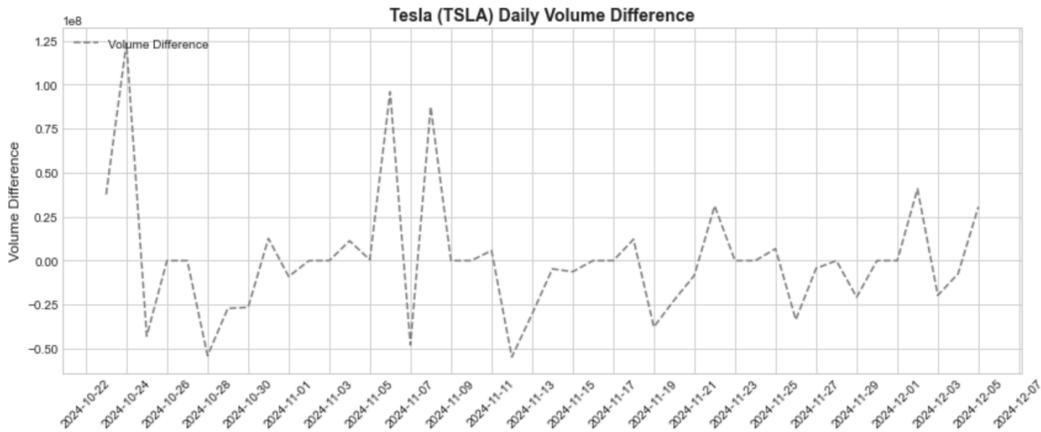

## 5. Conclusion

This study focuses on identifying optimal trading volumes and expected price ranges to execute trades with minimal market impact, ensuring stability. Our initial efforts are centered on building and testing the model using data from a single stock, allowing us to fine-tune parameters, assess performance, and establish robustness in a controlled environment.

Looking ahead, we aim to extend the study to a portfolio of stocks, enabling a comprehensive analysis of the model's scalability and applicability across diverse assets. This expansion would allow us to explore inter-asset correlations, evaluate risk-adjusted returns, and refine trading strategies for multi-asset portfolios.

The methodology developed in this study supports both limit order execution, where pre-specified prices and volumes are determined a day in advance, and intraday trading using real-time price ranges and trading volumes. The latter is particularly suited for short-term trading timelines, where market dynamics evolve rapidly.

Additionally, by comparing predicted trading prices and volumes with actual market data, we can calculate slippage—a key metric for evaluating intraday trading performance. This analysis not only provides insights into the efficiency of trade execution but also offers valuable strategy design.

## 6. Acknowledgments

Gratitude is extended to Professor Naftali Cohen for his invaluable support and insightful suggestions throughout this project. His guidance has been instrumental in shaping the direction of this work and enhancing its overall quality.

# Appendix

*Appendix 1: Data Sample for Tesla Inc.*

| Datetime | Open | High | Low | Close | Adj Close | Volume |
|---|---|---|---|---|---|---|
| 2024-10-22 09:30:00 | 217.0635986 | 217.2899933 | 216.2299957 | 216.2299957 | 216.2299957 | 2144156 |
| 2024-10-22 09:31:00 | 216.2299957 | 216.8099976 | 215.8500061 | 216.7962036 | 216.7962036 | 658621 |
| 2024-10-22 09:32:00 | 216.6199951 | 216.8300018 | 215.8399963 | 216.1741943 | 216.1741943 | 416735 |
| 2024-10-22 09:33:00 | 216.25 | 216.6298981 | 216.0601044 | 216.371994 | 216.371994 | 291416 |
| 2024-10-22 09:34:00 | 216.3999939 | 216.8800049 | 216.3800049 | 216.6100006 | 216.6100006 | 279045 |
| 2024-10-22 09:35:00 | 216.6549988 | 217.3800049 | 216.3000946 | 217.1300049 | 217.1300049 | 367297 |

*Appendix 2: 5-Minute Data Aggregation*

| Datetime | High | Low | Open | Close | Range | Range_Up | Range_Down |
|---|---|---|---|---|---|---|---|
| 2024-10-31 13:30:00+00:00 | 259.750000 | 256.549988 | 258.019989 | 258.089996 | 3.200012 | 1.730011 | 1.470001 |
| 2024-10-31 13:35:00+00:00 | 258.250000 | 256.049988 | 258.075012 | 256.600006 | 2.200012 | 0.174988 | 2.025024 |
| 2024-10-31 13:40:00+00:00 | 258.290009 | 256.414612 | 256.640015 | 258.179993 | 1.875397 | 1.649994 | 0.225403 |

*Appendix 3: Merged Subset for Distribution Identification*

| | Datetime | Volume | Range |
|---|---|---|---|
| 0 | 2024-10-31 13:30:00+00:00 | 3834876 | 3.200012 |
| 1 | 2024-10-31 13:31:00+00:00 | 442298 | 3.200012 |
| 2 | 2024-10-31 13:32:00+00:00 | 336652 | 3.200012 |
| 3 | 2024-10-31 13:33:00+00:00 | 377530 | 3.200012 |
| 4 | 2024-10-31 13:34:00+00:00 | 300394 | 3.200012 |

*Appendix 4: Data Aggregation at a daily level for EWMA calculation*

| | Datetime | High | Low | Open | Close | Volume | Range |
|---|---|---|---|---|---|---|---|
| 0 | 2024-10-31 00:00:00+00:00 | 259.750000 | 249.279999 | 258.019989 | 249.779907 | 153744.804688 | 10.470001 |
| 1 | 2024-11-01 00:00:00+00:00 | 254.000000 | 246.630005 | 252.042999 | 248.850006 | 138079.025974 | 7.369995 |

# Code

The reference code, which can be used for replication purposes, is available at the following link

# References

1. Adewuyi, Adejumo Wahab. "Modelling Stock Prices with Exponential Weighted Moving Average (EWMA)." *SCIRP*, Scientific Research Publishing, 5 Feb. 2016, www.scirp.org/journal/paperinformation?paperid=63814.
2. Andrieu, C., Doucet, A., and Holenstein, R. (2011). "Particle Markov chain Monte Carlo (with discussion)." J. Royal Statist. Society Series B, 72 (2): 269–342.
3. Case Study, Columbia University (Ali Hirsa), "Volume Volatility Order Slicing using MCMC", Ali Hirsa
4. Hachicha, A., & Hachicha, F. (2020, July 6). *Analysis of the Bitcoin stock market indexes using comparative study of two models SV with MCMC algorithm - Review of Quantitative Finance and Accounting*. SpringerLink. https://link.springer.com/article/10.1007/s11156-020-00905-w
5. Medium. (2023, April 15). *Metropolis-Hastings algorithm MCMC from scratch in Python*. https://exowanderer.medium.com/metropolis-hastings-mcmc-from-scratch-in-python-c21e53c485b7
6. Yahoo Finance. *Tesla, Inc. (TSLA) stock price, news, Quote & History*. Yahoo! Finance. https://finance.yahoo.com/quote/TSLA/